\documentclass[preprint,onecolumn,amsmath,letterpaper,amssymb,prl,floatfix]{revtex4-1}

\usepackage{graphicx}
\usepackage[font=small, justification=centerlast]{caption}
\usepackage{color}

\begin{document}

\title{Phenomenological and microscopic theories for catch bonds}

\author{Shaon Chakrabarti$^{1,2}$, Michael Hinczewski$^3$, and D. Thirumalai$^4$}
\affiliation{$^1$ Department of Biostatistics, Harvard T. H. Chan School of Public Health, Boston, MA 02115\\
$^2$ Department of Biostatistics and Computational Biology, Dana-Farber Cancer Institute, Boston, MA 02215\\
$^3$ Department of Physics, Case Western Reserve University, OH 44106 \\
$^4$ Biophysics Program, Institute For Physical Science and Technology, University of Maryland, College  Park, MD 20742
  }
  
\begin{abstract}
  Lifetimes of bound states of protein complexes or biomolecule folded
  states typically decrease when subject to mechanical force.
  However, a plethora of biological systems exhibit the
  counter-intuitive phenomenon of catch bonding, where non-covalent
  bonds become stronger under externally applied forces. The quest to
  understand the origin of catch-bond behavior has lead to the
  development of phenomenological and microscopic theories that can
  quantitatively recapitulate experimental data. Here, we assess the
  successes and limitations of such theories in explaining
  experimental data. The most widely applied approach is a
  phenomenological two-state model, which fits all of the available
  data on a variety of complexes: actomyosin, kinetochore-microtubule,
  selectin-ligand, and cadherin-catenin binding to filamentous
  actin. With a primary focus on the selectin family of cell-adhesion
  complexes, we discuss the positives and negatives of
  phenomenological models and the importance of evaluating the
  physical relevance of fitting parameters.  We describe a microscopic
  theory for selectins, which provides a structural basis for catch
  bonds and predicts a crucial allosteric role for residues
  Asn82--Glu88. We emphasize the need for new theories and simulations
  that can mimic experimental conditions, given the
  complex response of cell adhesion complexes to force and their
  potential role in a variety of biological contexts.
\end{abstract}

\maketitle
\def\s{\rule{0in}{0.28in}}

\renewcommand{\baselinestretch}{2}
\small\normalsize

\section{Introduction}

\begin{figure}
 \begin{center}
\includegraphics[width=5.5 in]{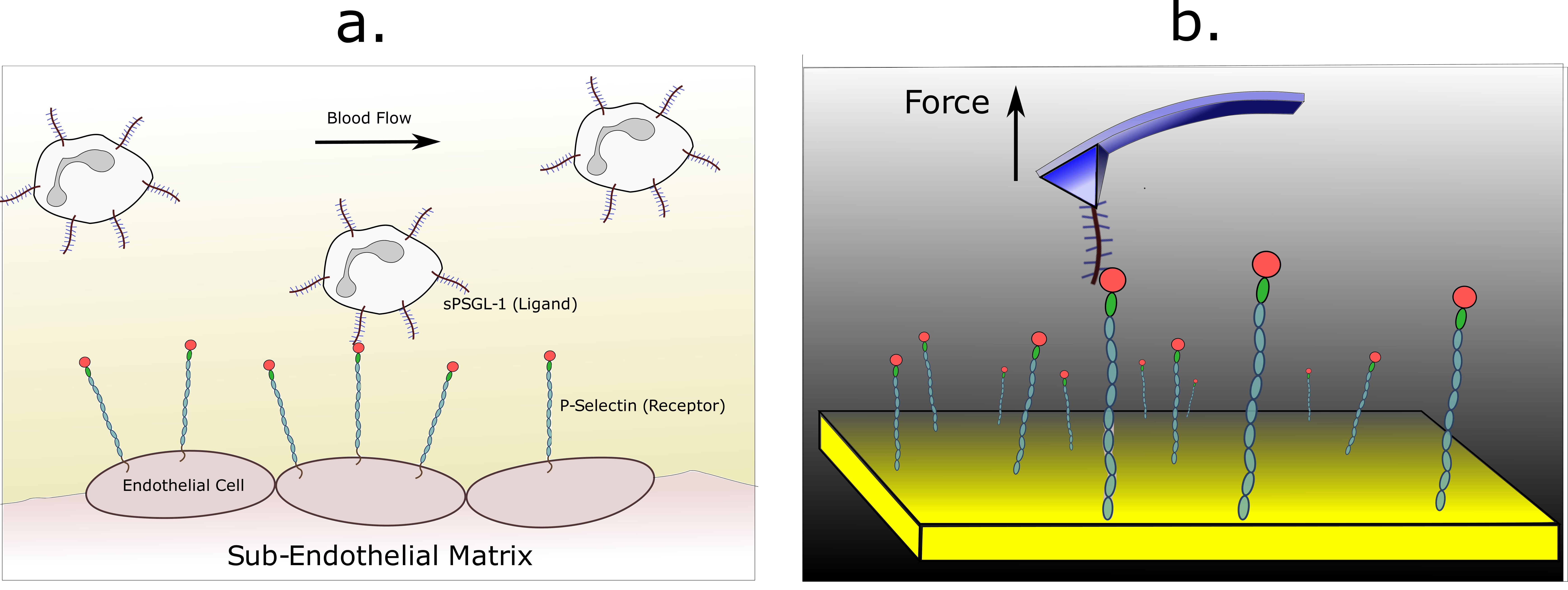}
\end{center}
\renewcommand{\baselinestretch}{1}
\small\normalsize
\caption[Probing P-selectin and PSGL-1 interactions] {Probing
  receptor-ligand interactions. (a) The cartoon shows ligands on
  leukocytes in the blood flow interacting with receptors on the
  endothelial cells. This interaction leads to the phenomenon of white
  blood cell rolling, and is the first step of a signaling cascade
  that ultimately leads to leukocyte localization at injured sites and
  wound healing. (b) Probing the receptor-ligand interaction at the
  single molecule level using an atomic force microscope
  (AFM). }\label{celladh}
\end{figure}

For complex multicellular organisms to function, individual cells need
mechanisms to bind to each other and to the extracellular matrix.
This is accomplished through specialized molecules on the surfaces of
cells known as adhesion proteins~\cite{berrier_cell-matrix_2007,
  gumbiner_cell_1996}.  Beyond their role as the essential mortar of
tissue architecture, these proteins are involved in signaling,
cellular movement, and tissue repair.  For example, the adhesion of
leukocytes to the endothelial cells of the blood vessel is a vital
step in rolling and capture of blood cells (Fig~\ref{celladh}a),
ultimately leading to wound
healing~\cite{ley_getting_2007,vestweber_mechanisms_1999,mcever_perspectives_1997}.
Viruses and bacteria utilize these molecules to establish initial
attachments with host-cell
receptors~\cite{marsh_virus_2006,pizarro-cerda_bacterial_2006}. The
general importance of cell adhesion complexes is evident from the fact
that many diseases are caused by the malfunctioning or faulty
expression of the proteins---for instance, the family of leukocyte
adhesive deficiency (LAD) diseases in humans
\cite{parsons_cell_2010,symposium_cell_2008,anderson_leukocyte_1987}.

In the process of executing their functions, cell adhesion complexes
are typically subject to fluid flows, which result in shear
stresses. Though these fluid flows sometimes impede the formation of
protein complexes, in many cases the generated shear forces are of
crucial functional importance. For instance, selectin and integrin
activation, leading to enhanced ligand binding, is only possible in the
presence of such shear flows
\cite{alon_cells_2008,alon_force_2007}. Biological function can also
be induced by other kinds of mechanical forces, such as those arising
from the coupling of focal adhesions to the cytoskeleton
\cite{burridge_focal_1996,wozniak_focal_2004}. Under stress, molecules
undergo conformational changes, triggering biophysical, biochemical,
and gene regulatory responses that have been, and still are, subjects
of intense research~\cite{davies_flow-mediated_1995,
  traub_laminar_1998}.

\begin{figure}
 \begin{center}
\includegraphics[width=6.5 in]{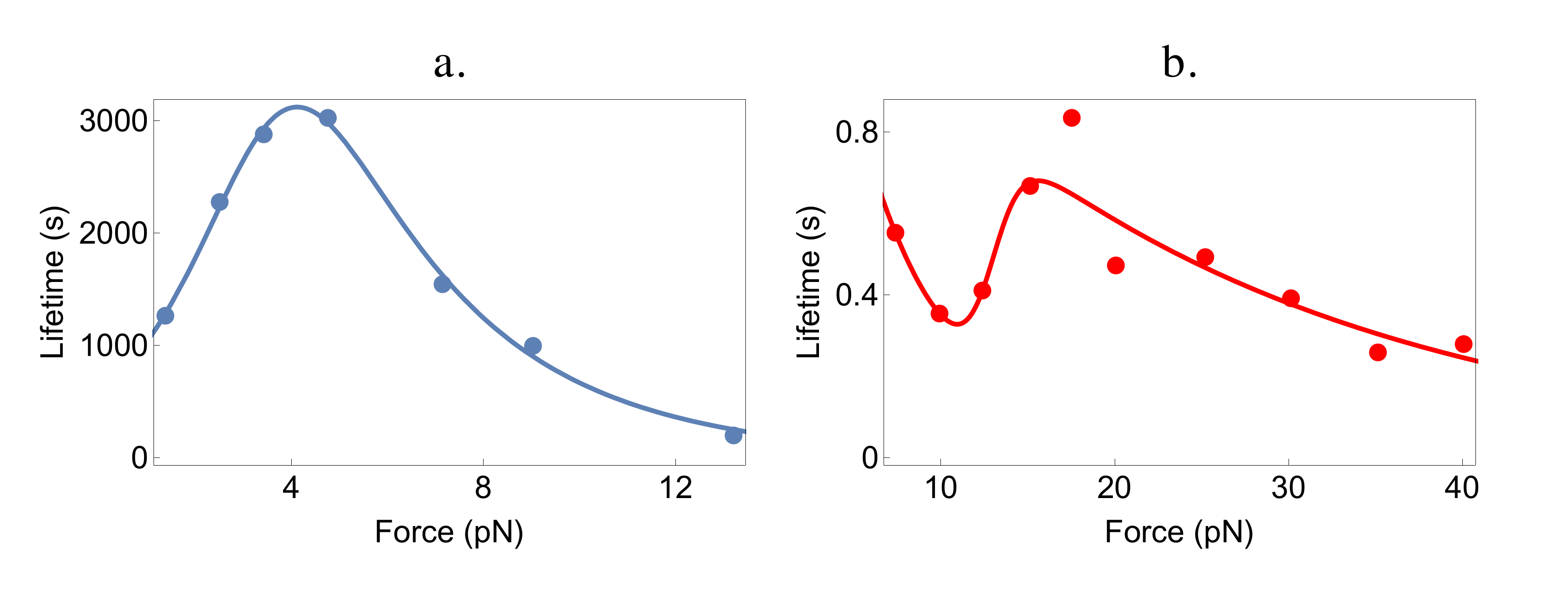}
\end{center}
\renewcommand{\baselinestretch}{1}
\small\normalsize
\caption[Biphasic and Triphasic (slip-catch-slip) experimental data]
{Catch bond data from (a) Kinetochore-microtubules
  \cite{akiyoshi_tension_2010} and (b) sulfatase-glycosaminoglycans
  \cite{harder_catch_2015}. The filled circles are experimental data
  while the lines are fits using the general two-state
  model. Evidently, both biphasic and triphasic lifetime behavior can
  be explained by this model. }\label{exp}
\end{figure}

One expects that a force acting on a protein assembly should decrease
its lifetime, the mean length of time the complex remains intact
before rupture.  This is indeed the experimental observation in a
multitude of cases. Such behavior, described phenomenologically by
Zhurkov~\cite{Zhurkov65} and Bell ~\cite{bell_models_1978}, is the
defining characteristic of a ``slip bond''. However, the response of
certain complexes to mechanical force exhibits a surprisingly
counterintuitive phenomenon. Lifetimes increase over a range of low
force values, so-called ``catch bond''
behavior~\cite{dembo_reaction-limited_1988}, while at high forces the
lifetimes decrease (Fig.~\ref{exp}a).  The non-monotonic response of a
variety of protein-complexes has attracted a great deal of attention
thanks to the ability to observe them directly in single molecule
pulling experiments \cite{Zoldak13COSB}
(Fig.~\ref{celladh}b). However, in retrospect, the existence of
catch-bonds was already evident in early experiments by Greig and
Brooks, who discovered that agglutination of human red blood cells,
using the lectin concanavalin A, increased under
shear~\cite{greig_shear-induced_1979}. Although not interpreted in
terms of catch bonds, their data showed lower rates of unbinding with
increasing force on the complex. Direct evidence for catch bonds in a
wide variety of cell adhesion complexes has come from flow, atomic
force microscopy (AFM), biomembrane force probe (BFP) and optical
tweezer experiments in the last decade \cite{thomas_bacterial_2002,
  marshall_direct_2003,
  kong_demonstration_2009,liu_accumulation_2014,Buckley14Science},
along with examples from other load-bearing cellular complexes like
actomyosin bonds~\cite{guo_mechanics_2006} and microtubule-kinetochore
attachments~\cite{akiyoshi_tension_2010}.  A number of articles have
reviewed these results (see Refs.~
\cite{sokurenko_catch2008,mcever_rolling_2010,rakshit_biomechanics_2014,liu_molecular_2015}). The
interested reader should consult these articles for details of
experimental methodologies and a wider overview of the kind of systems
where catch bonds have been discovered.

In this perspective we investigate the basic principles of some of the
commonly used catch bond models critically. The successes and
limitations of the theories are pointed out. In the process, we
highlight the need for theories that account for the structural
transitions of protein complexes subject to force. This is critically
necessary because only by developing such theories testable
predictions can be made. To date there is only one microscopic theory
\cite{chakrabarti_2014}, applicable to a class of cell-adhesion
complexes, that satisfies this criterion.  We believe that progress in
understanding the role of catch bonds under {\it in vivo} conditions
can only be made by creating suitable theories with predictive power.

\renewcommand{\baselinestretch}{2}
\small\normalsize

\section{Phenomenological theories}

{\bf The two-state model:}

\begin{figure}
 \begin{center}
\includegraphics[width=2.0 in]{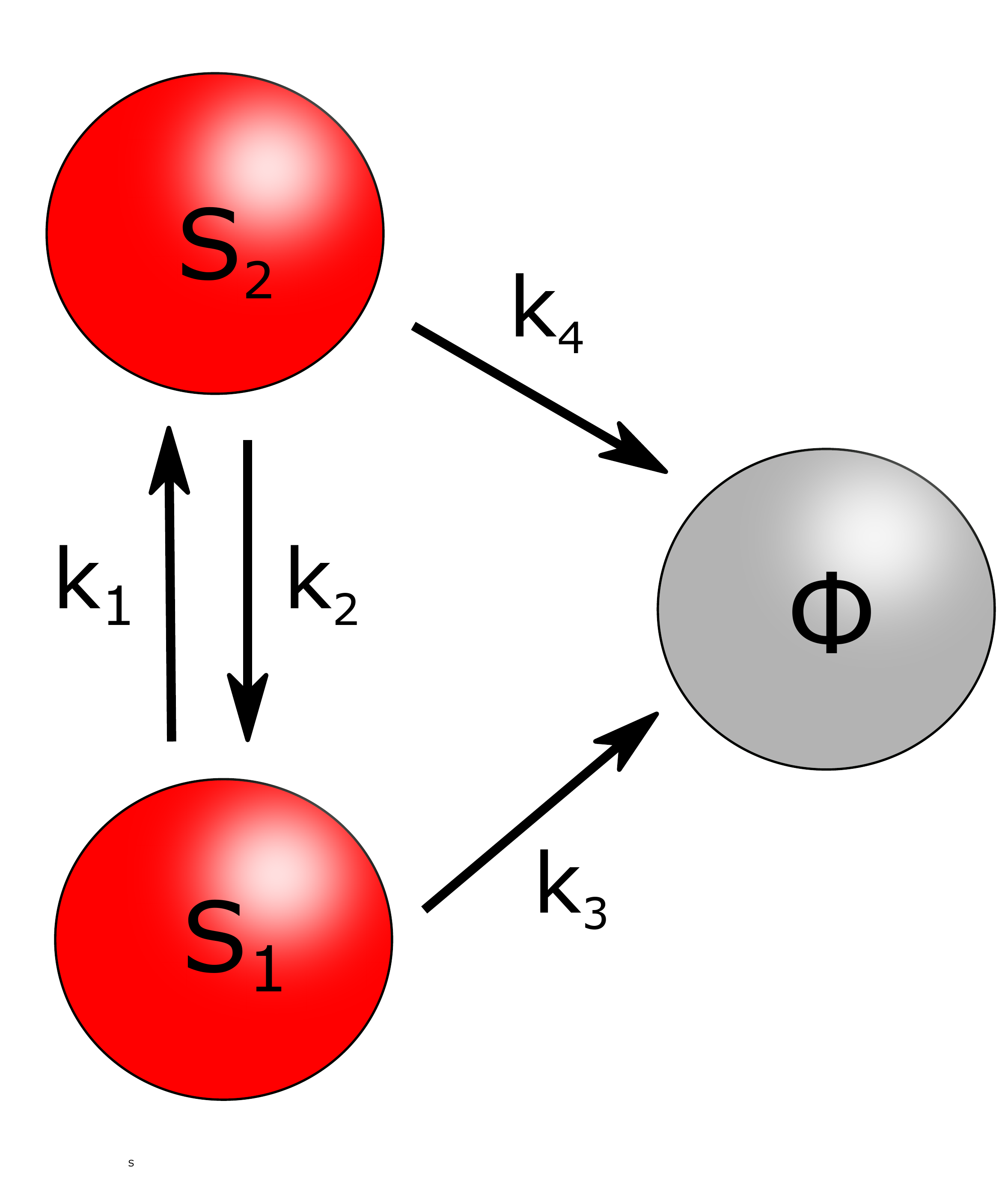}
\end{center}
\renewcommand{\baselinestretch}{1}
\small\normalsize
\caption{The two-state model for catch bonds. There are two
  protein-ligand bound states $S_1$ and $S_2$, which can interconvert
  with rates $k_1$ and $k_2$. From each state, the complex can
  disassociate to form $\Phi$ with characteristic rates $k_3$,
  $k_4$. All the rates depend on the external force in a Bell-like
  fashion, as described in the text. }\label{two}
\end{figure}

A theoretical explanation for catch-bonds at the single molecule level
was provided by Barsegov and Thirumalai (BT)
\cite{barsegov_dynamics_2005}, inspired by experiments on
forced-unbinding of complexes of P-selectin with ligands. The
essential idea is that the protein-ligand complex can exist in two
bound states $S_1$ and $S_2$ as depicted pictorially
(Fig.~\ref{two}). The model in \cite{evans_mechanical_2004} is often
considered to be a two-state model for catch bonds. However, it is
worth emphasizing that there are key differences between the
approaches in \cite{evans_mechanical_2004} and
\cite{barsegov_dynamics_2005}. In the former it was assumed that the
two states of the complex interconvert rapidly, thus restricting the
application of the model to the analysis of only a few
experiments. The complete solution of the simple two-state model was
provided by BT, which can be used to study catch bonds in all systems.

\renewcommand{\baselinestretch}{2}
\small\normalsize

The free energy barrier between the two states determines how fast
they interconvert (with rates $k_{10}$ and $k_{20}$ at zero
force). The protein-ligand complex could dissociate to a state $\Phi$,
with characteristic zero-force dissociation rates $k_{30}$ and
$k_{40}$ for $S_1$ and $S_2$ respectively. If $k_{40} > k_{30}$,
dissociation from $S_2$ would be easier than dissociation from $S_1$,
implying that the energy barrier along $S_2 \rightarrow \Phi$ is lower
than that along $S_1 \rightarrow \Phi$. To describe the effect of an
external force ($F$) on this energy landscape, each of the four rates
was assumed to vary with $F$ according to the Bell equations,
$k_i(F)=k_{i0} \, e^{d_i F}$ ($i=1,2,3,4$). The reason the two-state
model could produce non-monotonic lifetimes as a function of $F$ is
evident from the following scenario.  If initially (at zero $F$) a
large fraction of the protein-ligand population is in $S_2$, most
dissociation of the complex would occur from $S_2$, making the average
lifetime small. With an increase in $F$, the force-stabilized $S_1$
starts becoming more populated, thus leading to dissociation events
from both $S_1$ and $S_2$. This would naturally result in larger
average lifetimes, thus giving the catch bond regime.  Beyond a
certain critical force $F_c$, the bound protein-ligand population
would be almost entirely in $S_1$, and the system returns to a slip
bond regime, characterized by a mean lifetime decaying with $F$ with
rate constant $d_1$. The two-state model was successfully used to
explain catch and slip bond data from P-selectin and its ligands
\cite{barsegov_dynamics_2005,Barsegov06JPCB}, therefore providing an
important and basic physical understanding of the apparently strange
catch bond phenomenon.

The two-state model has been subsequently used to explain catch-slip
data from a number of biological systems like the bacterial FimH
adhesive protein \cite{thomas_catch-bond_2006,Pereverzev2009},
kinetochore-microtubule attachments \cite{akiyoshi_tension_2010}, cell
surface sulfatase and glycosaminoglycan interactions
\cite{harder_catch_2015} and cadherin-catenin interactions
\cite{Buckley14Science}. Among all these experiments, the work by
Akiyoshi {\it et al.} on kinetochore-microtubule attachments
\cite{akiyoshi_tension_2010} is a remarkable validation of the
two-state model. The four force-dependent rates $k_1(F)$ through
$k_4(F)$ were measured directly in their experiment, which were then
used in the two-state model with no free parameters to reproduce the
experimental catch-slip lifetime data.  Catch bond behavior in the
cadherin-catenin/F-actin complex demonstrated in
Ref.~\cite{Buckley14Science} is also noteworthy.  The complex with
F-actin, known to readily form {\it in vivo}, can only be
reconstituted {\it in vitro} in the presence of force.  This suggests
that the {\it in vivo} complex is likely under tension.  The
experimental force-dependent mean lifetimes and the survival
probability of the the minimal complex comprising cadherin-catenin and
F-actin were analyzed quantitatively \cite{Buckley14Science} using the
two-state model exactly as formulated by BT, who analyzed the data on
the selectin-ligand complex by assuming equilibrium between $S_1$ and
$S_2$ without invoking any force history on the initial population
distribution.

{\bf Slip-catch-slip transition:} The BT model also predicts that in
principle it is possible to observe a decrease in the lifetime of a
bond (slip bond) at $F \le F_{min}$ followed by an increase in the
lifetime (catch bond) in the intermediate force regime,
$F_{min} < F \le F_c$, and finally a decrease in the lifetime at
$F \ge F_c$. Such a scenario is possible when at small forces
($F \le F_{min}$) the force-stabilized state cannot be populated
sufficiently so that unbinding occurs mainly from the weakly bound
state $S_1$. This would lead to an initial regime of conventional slip
bond behavior. The predicted triphasic (slip-catch-slip) behavior
should be generic although it appears that in many cases $F_{min}$
could be very small, thus preventing detection of the initial slip
bond behavior. However, this triphasic behavior has been observed in
an insightful experiment probing cell surface sulfatase and
glycosaminoglycan interactions \cite{harder_catch_2015}
(Fig.~\ref{exp}b), and also in an experiment on the von Willebrand
factor \cite{Kim09Nature}. Although not analyzed in terms of triphasic
behavior, it appears that force effects on the von Willebrand factor
seem to be in accord with this slip-catch-slip scenario.

{\bf Effective 1-D models:}

Besides the two-state model, a variety of effectively 1-D models have
been proposed and used to analyze catch bond data
\cite{Pereverzev2005,fei_dynamic_2006}. The most widely used among
these is the ``one-state, two-pathway'' model
\cite{Pereverzev2005}. Based on the original models proposed in
\cite{denis_bartolo_dynamic_2002}, this model posits one
protein-ligand bound state instead of two, and allows for bond rupture
via two different pathways. The two pathways have barriers of
different heights which the bound state complex must overcome in order
to dissociate. Under different force conditions a varying fraction of
the bound state population escape via the two pathways, thereby giving
rise to catch bond phenomena. Unlike the two-state model, which has
experimental validation (especially in kinetochore-microtubule
complexes \cite{akiyoshi_tension_2010}), this model has not yet been
shown to have any direct experimental significance.

{\bf Limitations of the phenomenological theories:}

Although the two-state model has been used to recapitulate an
impressive range of experimental data sets, a major limitation of the
model is that it does not provide a structural explanation for the
origin of catch bonds.  In this picture, force ($F$) is coupled (in a
Zhurkov-Bell exponential manner) to the distances $d_1$, $d_2$, $d_3$
and $d_4$, which are meant to represent transition state (TS)
distances. If stretching by force exceeds the TS distance then the
bound state is destabilized.  However, since the actual protein-ligand
energy landscape is multi-dimensional, comprising the coordinates of
all the atoms, the TS transition states are merely projections along
the force direction. Without any structural knowledge of the complex
landscape or assessing the adequacy of such projections
\cite{morrison_compaction_2011}, it becomes difficult to extract any meaningful
information from a knowledge of these distances alone. Though if the
extracted distances are physically reasonable it adds to the
credibility of the phenomenological two-state theory.

The effective 1-D models have fewer free parameters than the two-state
model, but nonetheless their efficacy is tarnished by their inability
to produce physically reasonable parameters when analyzing
experimental data. For example, the one-state, two-pathway model
produces a non-physical negative transition state distance when used
to analyze catch bond data \cite{Pereverzev2005}. This immediately
suggests that these distances are projections, and provides no
theoretical basis to reconstruct the actual transition state
distances, assuming that this notion is even appropriate.  The failure
of effective 1-D models has recently been highlighted by Zhuravlev
{\it et al.} \cite{Zhuravlev16PNAS}, who showed using a very general
theory that the energy landscape of systems exhibiting catch bond
behavior has to be strongly multidimensional, making any effective 1-D
theory inadequate.

In light of the arguments given above it is clear that one has to
create theories that capture the crucial structural features of the
protein-ligand complexes. We now discuss a microscopic theory devised
with an eye towards understanding the structural origins of catch
bonds. Since structural models are by definition more limited in their
scope and applicable to specific biological systems, the ensuing
discussion will be much less general than the previous one. We will
focus on P-selectin and its ligands, which along with L-selectin is
the only system for which microscopic models have been used to analyze
data \cite{chakrabarti_2014}.  .

\section{Microscopic models for the unbinding of selectin-ligand
  complexes}

\begin{figure}
 \begin{center}
\includegraphics[width=4.0 in]{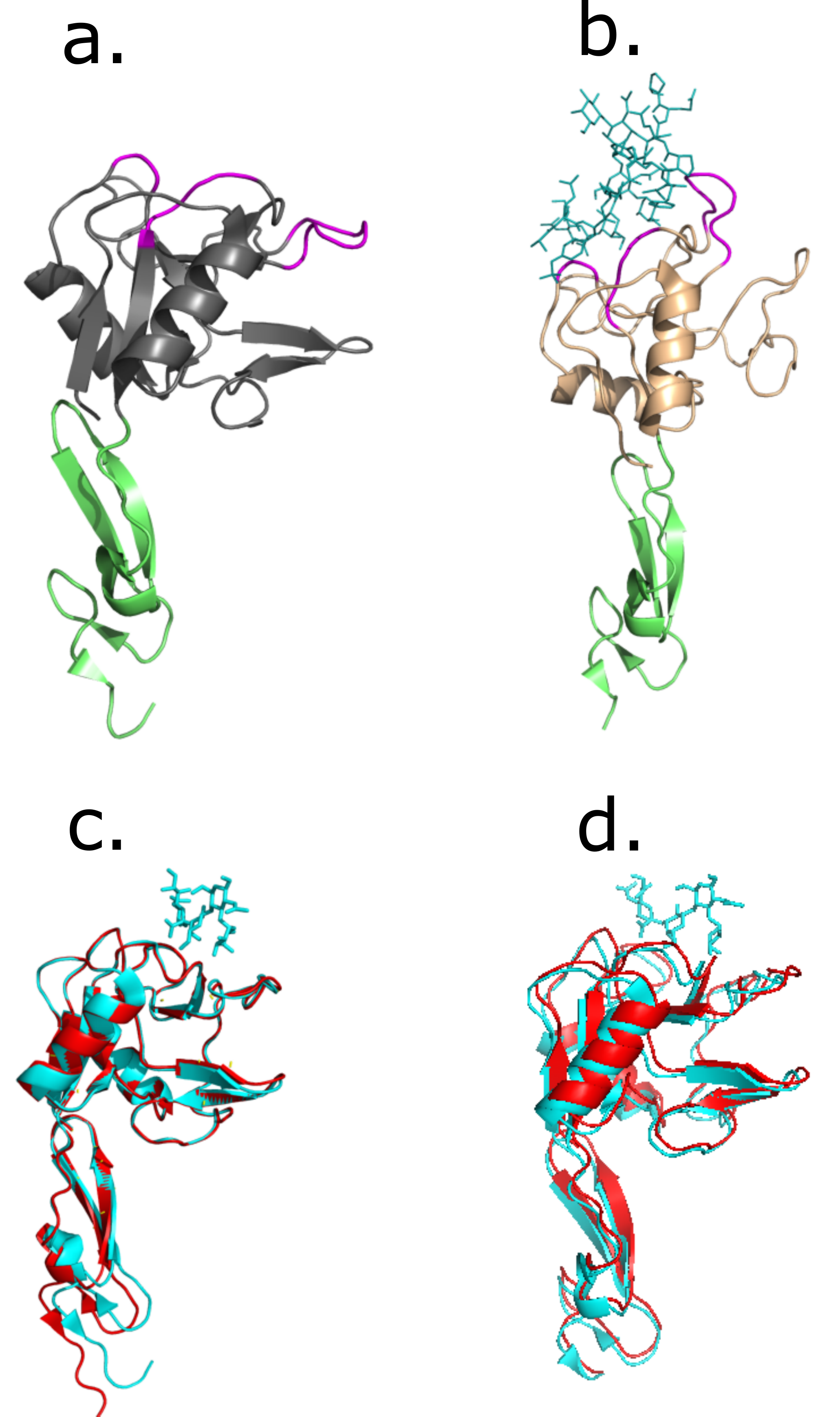}
\end{center}
\renewcommand{\baselinestretch}{1}
\small\normalsize
\caption[P and E-selectin structures with and without ligands] {Selectin
  structures with and without ligands. (a) P-selectin in the bent
  state (PDB ID 1G1Q). (b) P-selectin complexed with ligand in
  the extended state (PDB ID 1G1S). (c) Aligned P-selectin structures in
  the bent state, with (1G1R, blue) and without (1G1Q, red)
  ligand. (d) Aligned E-selectin structures in the bent state, with
  (1G1T, blue) and without (1ESL, red) ligand.}\label{pselstruc}
\end{figure}

{\bf Insights from experiments and crystal structures:} The idea is to
create an effective multi (at least two) dimensional energy landscape
that can be justifiably obtained from known structures of selectin
complexes. Key ingredients for a microscopic theory can be deduced by
analyzing experiments that provide both biochemical and structural
data for selectins \cite{somers_insights_2000,phan2006}. The
structures of a number of selectin complexes are shown in
Fig.~\ref{pselstruc}, both with and without
ligands. Fig.~\ref{pselstruc} a shows P-selectin in the ``bent'' or
``flexed'' state, while Fig.~\ref{pselstruc}b shows the same receptor
in the ``extended'' state. These are the only two states that have
been crystallized in the selectin family of receptors. The green
domain in both figures is the EGF domain, while gray/beige represents
the lectin domain. The purple regions are the ligand binding domains
of the receptor. As is evident from the two figures, the angle between
the EGF and lectin domains defines whether the receptor is in the bent
or extended state. In Fig.~\ref{pselstruc}c and d, the bent states of
P- and E-selectin are shown, with and without a ligand. Clearly, the
structure of the bent state does not really change with or without the
ligand. Fig.~\ref{pselstruc} as a whole, suggests that ligands can
bind the selectin receptor either in the bent state or in the extended
state.
\renewcommand{\baselinestretch}{2} \small\normalsize In addition,
mutation experiments provide evidence regarding the lifetime of the
ligands in the two conformational states of the selectin receptor. In
a beautiful experiment, Phan {\it et al.}~\cite{phan2006} created an
extra carbohydrate region (glycan) at the interface between the lectin
and EGF domains of P-selectin. The glycan domain acted as a wedge to
pry the lectin and EGF domain apart, forcing them to adopt only the
extended conformation. The lifetime of a ligand was then measured for
the mutant, and compared to the lifetime of the wild type, which
lacked the glycan wedge. Surprisingly, the lifetime of the mutant was
larger, indicating that the ligand bound the receptor more tightly in
the extended state compared to the bent state.

\begin{figure}
 \begin{center}
\includegraphics[width=4.0 in]{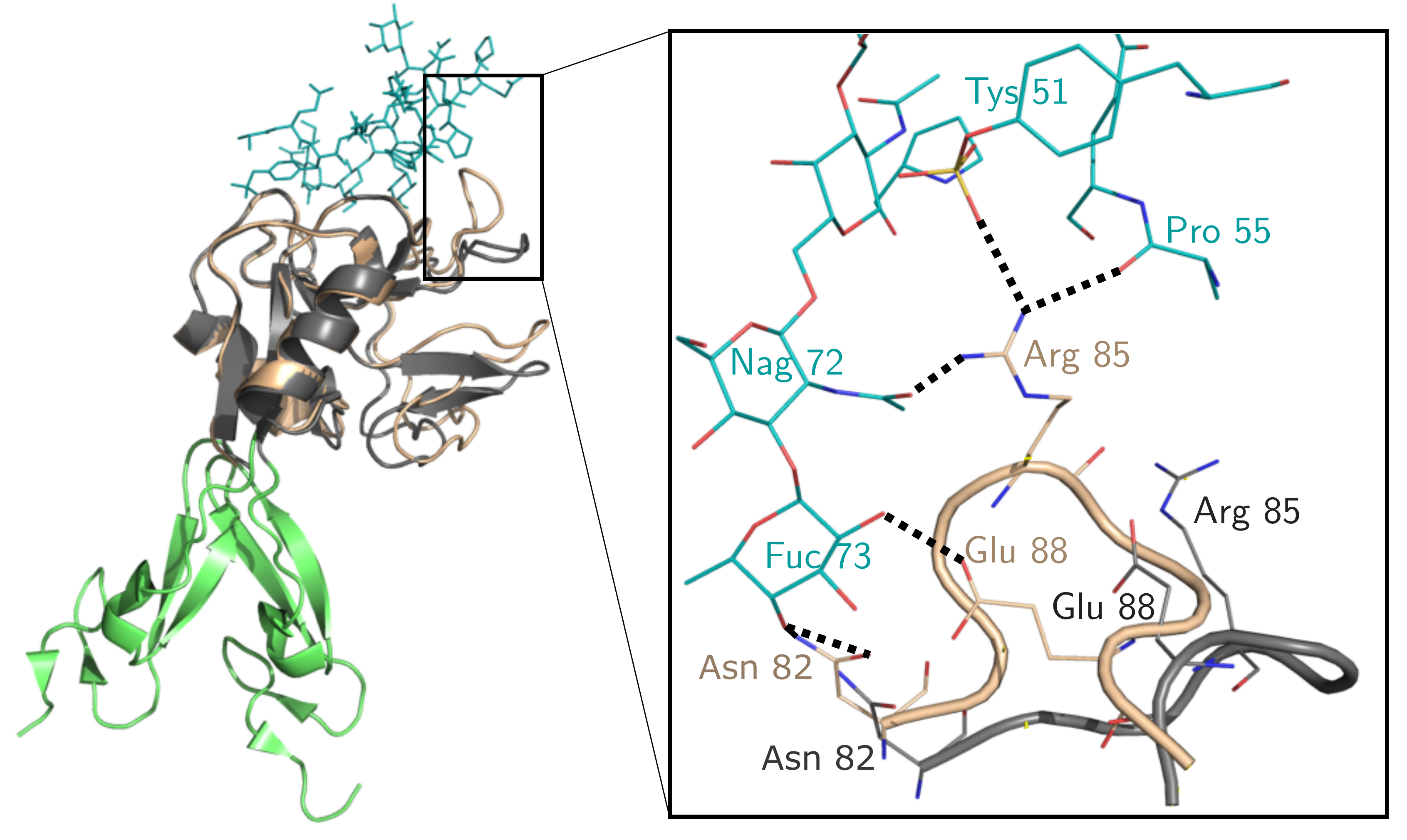}
\end{center}
\renewcommand{\baselinestretch}{1}
\small\normalsize
\caption {Close-up of the receptor-ligand interactions of P-selectin in the bent (dark grey) and extended (beige) states. In the extended state, residues in the loop Asn82-Glu88 create new hydrogen bonds (dashed lines) with the ligand (blue) that were not present in the bent state. }\label{zoom}
\end{figure}

A plausible reason for the larger ligand lifetime in the extended
state (and hence the catch bond phenomenon) can be inferred from an
analysis of the crystal structures shown in Fig.~\ref{pselstruc}. The
purple shaded loop in Fig.~\ref{pselstruc}a and b denotes the set of
residues between Asn82 and Glu88 that are part of the ligand-binding
lectin domain of P-selectin. As pointed out elsewhere
\cite{somers_insights_2000,springer_structural_2009}, there is a major
structural change in this loop, going from the bent
(Fig.~\ref{pselstruc} a, c and d) to the extended
(Fig.~\ref{pselstruc} b) conformations. Unlike the bent conformation
where the loop creates no contacts with the ligand, there are six
hydrogen bonds formed in the extended state (Fig.~\ref{zoom}). The
conformational changes have been suggested to arise due to allostery
\cite{waldron_transmission_2009}, which is supported by the
observation that the mutation A28H in the lectin domain, that is far
from either the ligand binding interface or the lectin-EGF interface,
can cause an increase in affinity for the ligand
\cite{waldron_transmission_2009}.
\renewcommand{\baselinestretch}{2} \small\normalsize In summary,
experimental evidence from biochemical studies and crystal structures
suggests that selectins can exist in (at least) two conformations---a
bent and an extended state. Both states can bind ligands, but
crucially, the lifetime is larger in the extended state. The larger
lifetime in the extended state could be due to structural changes in
the loop of residues Asn82-Glu88, which create extra contacts with the
ligand only in the extended state.

\subsection{Structure-based energy-landscape model predicts crucial
  role of Asn82-Glu88 loop and allostery in selectin catch bonds}

\begin{figure}
 \begin{center}
\includegraphics[width=4.0 in]{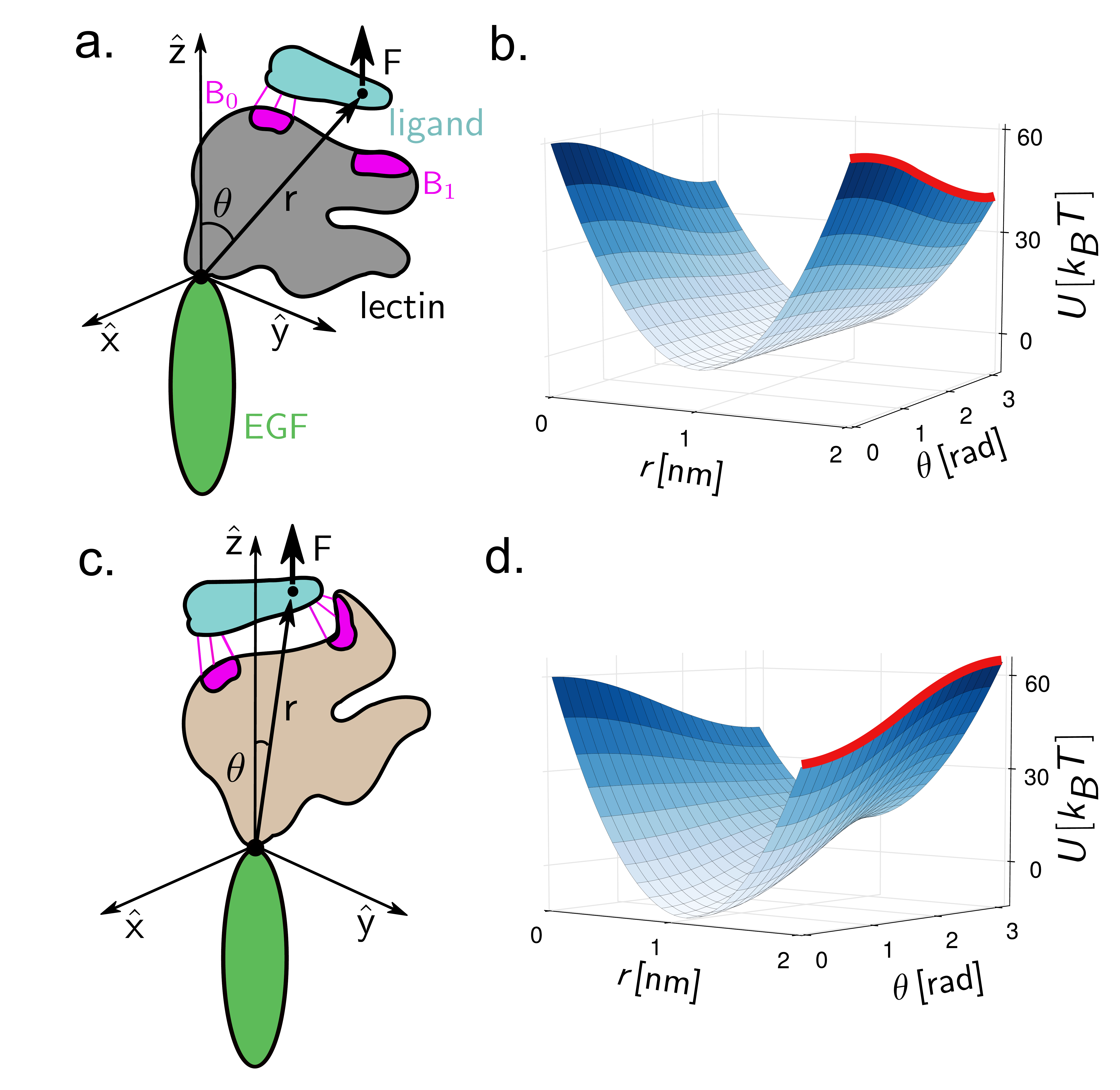}
\end{center}
\renewcommand{\baselinestretch}{1}
\small\normalsize
\caption {Structure-based microscopic model for catch bonds in selectins, developed by Chakrabarti
  {\it et al.} \cite{chakrabarti_2014}. a), c): The model, highlighting the key components. b), d): The energy landscape at zero and high forces respectively. The barrier to bond breaking is shown as a red line. }\label{model}
\end{figure}

The experimental results provide fundamental insights to the possible
origin of catch bond behavior in selectins, and are reminiscent of the
two-state model \cite{barsegov_dynamics_2005,Barsegov06JPCB} with the
extended and bent states serving as the two bound states $S_1$ and
$S_2$ discussed earlier in this review. However, to quantitatively
judge whether the shift of the loop region Asn82-Glu88 can indeed
explain P-selectin force-lifetime curves in single molecule
experiments \cite{marshall_direct_2003}, we created a microscopic
model (described in detail in Ref.~\cite{chakrabarti_2014}).  A
three-dimensional energy landscape (effectively two-dimensional due to
azimuthal symmetry in the Hamiltonian) was designed to mimic the
angle-dependent ligand detachment rate observed in experiments, and to
allow the external force to be incorporated in a manner similar to the
geometry in single molecule AFM experiments (Fig.~\ref{model}). Note
that an a priori Bell-like force dependence was not assumed in this
model, but emerged naturally only at large forces
\cite{chakrabarti_2014}. Mean first passage times within the energy
landscape were then calculated using a Fokker-Planck formalism, to
estimate the dissociation rates at various forces. Crucially, the four
free parameters of the model were directly associated with quantities
that can be measured, for instance the number of hydrogen bonds in the
Asn82-Glu88 loop or the size of the lectin domain. The mathematical
model provided very strong support to the idea that remodeling of the
Asn82-Glu88 loop causes catch-like behavior in P-selectin. It also
provided a concrete prediction for change in force-lifetime behavior
on mutating the sulfated tyrosine 51 on PSGL-1 (the P-selectin ligand)
to phenylalanine. This mutant PSGL-1 construct had been developed
earlier \cite{Xiao2012}, and hence an experimental validation of the
prediction is possible, and would be of much interest in the context
of the model.

Finally, we point out the limitations of the above model in its
current form: A correct theory for catch bonds must be able to explain
the full distribution of experimentally determined lifetimes in
addition to the average lifetime as functions of force. Since this
model was built keeping specifically selectins in mind (which exhibit
single-exponential lifetime distributions), it cannot be used to
understand catch bonds in systems that exhibit double-exponential
lifetimes (for example cadherin-catenin \cite{Buckley14Science}).
Even more complex behavior has been observed in E-selectin, where
there is a slip-catch-slip triphasic behavior of the force-lifetime
curve \cite{wayman_triphasic_2010}. This too cannot be explained by
the current model and might be important to consider in the future.
As noted earlier, triphasic behavior has also been seen in the
interactions of cell surface sulfatase and glycosaminoglycans
\cite{harder_catch_2015}, where the authors explained the data using
the two-state model as formulated by BT
\cite{barsegov_dynamics_2005}. Finally, we should point out that
unlike the only available microscopic theory, so far restricted to the
selectin family \cite{chakrabarti_2014}, the phenomenological model
can be used to quantitatively analyze all of the available data.
\renewcommand{\baselinestretch}{2}
\small\normalsize

{\bf Sliding-rebinding model}

A very different model has been proposed to explain catch bonds in a
variety of other adhesion complexes
\cite{lou_flow-enhanced_2006,lou_structure-based_2007}. This
``sliding-rebinding'' model was originally inspired by results of
steered molecular dynamics simulations on L-selectin
\cite{lou_flow-enhanced_2006} (albeit at unphysically large loading
rates), and proposes a radically different explanation for the
increase in lifetimes under force. The observation in the simulation,
which uses very large values of $F$ to observe rupture in very short
times, was that under external force, the ligand shifted its position
in the binding pocket of the lectin domain, thereby rupturing original
bonds and creating new bonds that stabilized the ligand.  A model was
proposed based on this observation, which can be summarized as
follows: $N$ pairs of pseudo-atoms represent the non-covalent bonds of
the zero force receptor-ligand complex, where $N$ can be any number
greater than zero. Under force, sliding results in a progressive
decrease in the number of interacting pairs while rebinding increases
the number of interacting pairs. Each event is given a defined rate of
formation/disruption (with Bell-like force dependence), and the final
set of rates is used in a kinetic Monte Carlo simulation to calculate
lifetimes of the receptor-ligand \cite{lou_structure-based_2007}.

The principle of sliding of protein domains to form new stabilizing
contacts seems to have experimental support
\cite{rakshit_ideal_2012,manibog_2014}. Using steered molecular
dynamics simulations of cadherin molecules in the presence of calcium
ions, Manibog {\it et al.} showed that the sliding of opposing
cadherins under force can cause formation of new hydrogen bonds
\cite{manibog_2014}. Based on their simulations they predicted that
reducing the calcium ion concentration would eliminate the
force-induced hydrogen bonds, which was then validated in AFM
experiments.  Though these results indeed support the basic idea of a
sliding-induced stabilization, the mathematical analysis carried out
based on the sliding-rebinding model highlights an inherent issue with
the model, that has not been addressed satisfactorily to date. Like in
the original papers \cite{lou_structure-based_2007}, Manibog {\it et
  al.}  observe an unphysically large (more than two orders of
magnitude) rebinding rate of interactions compared to the regular bond
formation rate. In addition, the model has features that are difficult
to justify on physical grounds and difficult to measure
experimentally---for instance, a force scale $f_0$ beyond which new
interactions are formed with unit probability. Without a more physical
justification of the extracted parameters of the model, it therefore
becomes difficult to judge the validity of the sliding-rebinding
model. Finally, the lifetime distributions predicted by the
sliding-rebinding model were not explored by the authors in the
cadherin study \cite{manibog_2014}. The experimental data clearly
suggests double-exponential lifetime distributions, and it is not yet
clear whether the sliding-rebinding model can produce similar results
for the parameters extracted.

For the particular case of catch bonds in selectins, it has been
pointed out before that steered molecular dynamics simulations and the
sliding-rebinding model fail to reproduce essential experimental
details \cite{waldron_transmission_2009}. To begin with, the atomistic
simulations on L-selectin where force was used to convert the bent
state to the extended state, could not reproduce the crucial
structural change in the Asn82--Glu88 loop seen in crystal structures
(Figure 1B in \cite{lou_flow-enhanced_2006}). This is hardly
surprising since the current force fields in all-atom simulations,
especially those for divalent ions, are not good enough to reproduce
major allosteric changes in protein domains. This is not an isolated
incident, and major differences between SMD predictions
\cite{puklin2006} and eventual crystal structures have been noted in
the case of $\alpha_5 \beta_3$ integrin as well
\cite{zhu_complete_2013}. In addition, it has been shown in recent
experiments that at high ramp rates of pulling, the catch bond
behavior of certain complexes completely disappears due to
non-equilibrium effects \cite{Sarangapani2011}. It is therefore
difficult to justify using steered molecular dynamics simulations
(which usually operate under ramp rates that are several orders of
magnitude larger than experimental rates) to investigate the origins
of catch bond behavior.  In light of all these limitations, steered
molecular dynamics simulation results cannot be taken seriously until
force-fields are improved and the simulations are carried out at
forces that are comparable to those observed in experiments.  Because
of these difficulties we surmise that the sliding-rebinding model,
while plausible, should be viewed as unphysical.
\begin{figure}
 \begin{center}
\includegraphics[width=5.2 in]{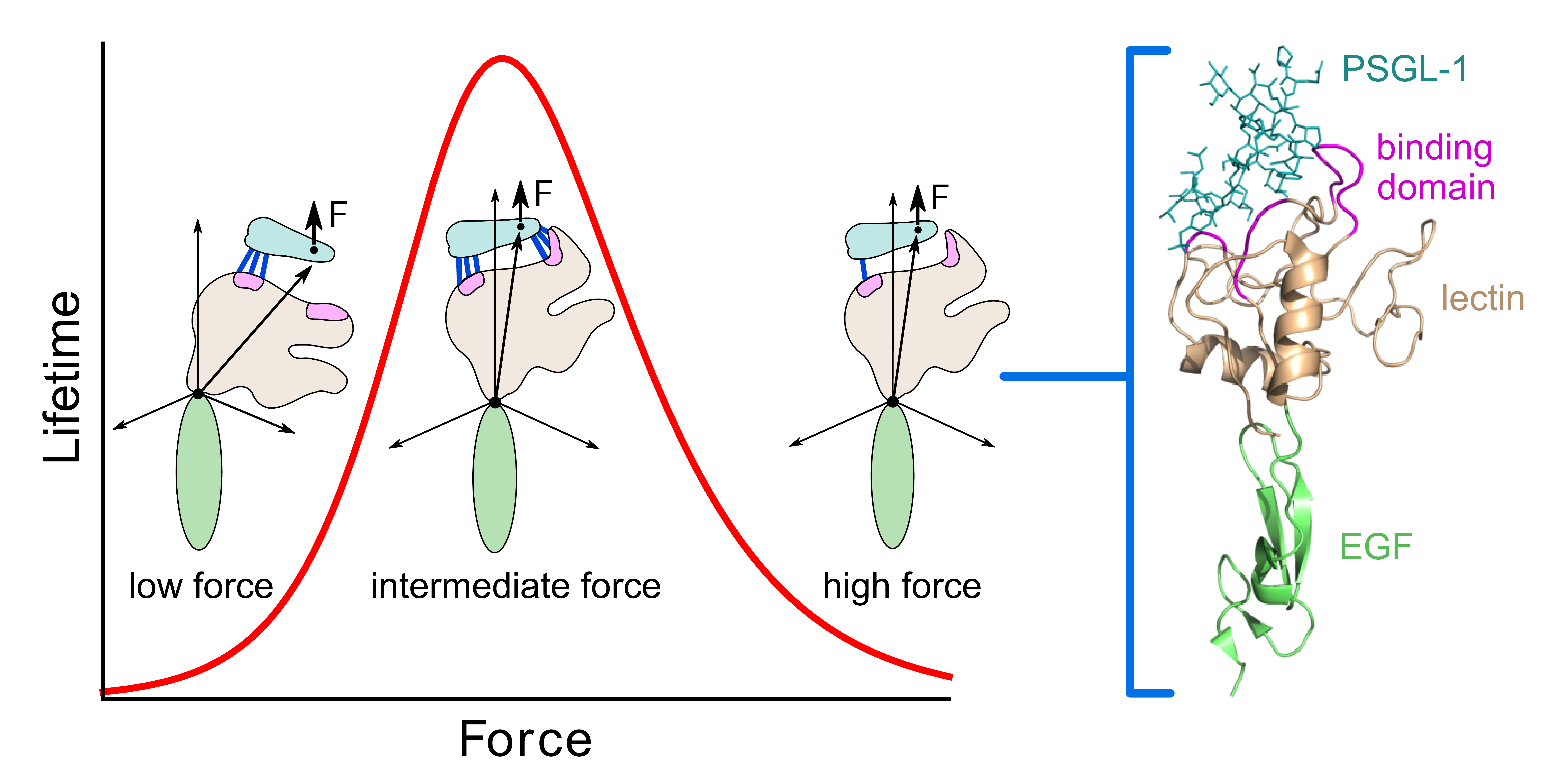}
\end{center}
\renewcommand{\baselinestretch}{1}
\small\normalsize
\caption[Structural basis of P-selectin and PSGL-1 catch bonds] { A
  summary of the predicted allosteric mechanism of catch bonds in
  P-selectin and PSGL-1. Allosteric changes coupled to the lectin
  domain rotation create extra hydrogen bonds between the receptor and
  ligand, causing catch bond behavior.}\label{summary}
\end{figure}

\section{Conclusions and future directions}

As the exciting field of mechanobiology hurtles into a new age of
experimental, theoretical and computational research, it is worthwhile
to pause for a moment and critically analyze the tools being currently
used to analyze experiments, in order to chart out the future path for
developing more informative theories. Here, we have explored some of
the theoretical ideas currently employed to analyze catch bond data
from experiments. We have highlighted the insights provided by
phenomenological theories over the last decade, yet at the same time
balanced it with discussions of their limitations. Phenomenological
theories of catch bonds must give way eventually to more detailed and
structure-based models, and we argue how one such model suggests a
clear structural mechanism for catch bonds in selectins
\cite{chakrabarti_2014} (Fig~\ref{summary}). Another recent work
explored kinetochore-microtubule catch-bonds based on an energy
landscape model \cite{sharma_distribution_2014}. Although experimental
data was not analyzed, it will be interesting if this model can shed
light on structural mechanisms in the future. A major theme of our
discussions has been interpretability and physical meaningfulness of
parameters extracted from mathematical models. In addition, we have
emphasized that detailed simulations must be performed under
conditions that mimic experimental forces and loading rates in order
to be trustworthy and relevant.  Given the observation of catch bonds
in diverse systems it is critical to create general theories, if
possible, in order to explain their origin and shed light on the way
nature uses them in executing cellular functions. In order to achieve
these goals the theories have to be critically evaluated on solid
physical grounds.

\begin{acknowledgments}
  This work was supported in part by grants from the National Science
  Foundation (Grant No. CHE 13-61946) and the National Institutes of
  Health (GM 089685).
\end{acknowledgments}

\bibliography{CatchBonds}

\end{document}